\documentclass[showpacs,preprintnumbers,amsmath,amssymb, prb, twocolumn]{revtex4-1}

\usepackage{stmaryrd}
\usepackage{txfonts}
\usepackage{amssymb}
\usepackage{mathrsfs}
\usepackage{graphicx}
\usepackage{dcolumn}
\usepackage{bm}
\usepackage{epsfig}
\usepackage{xcolor}

\begin{document}

\title{Fractional Flux Plateau in Magnetization Curve of Multicomponent Superconductor Loop}

\author{Zhao Huang\(^{1,2}\) and Xiao Hu\(^{1,2}\)}

\affiliation{\(^{1}\) {International Center for Materials Nanoarchitectonics (WPI-MANA), National Institute for Materials Science, Tsukuba 305-0044, Japan}
\\
\(^{2}\)Graduate School of Pure and Applied Sciences, University of Tsukuba, Tsukuba 305-8571, Japan}

\pacs{74.25.Uv, 74.20.Rp, 74.70.Xa, 74.20.De} 

\begin{abstract}
Time-reversal symmetry (TRS) may be broken in superconductors with three or more condensates interacting repulsively, yielding two degenerate states specified by chirality of gap functions. We consider a loop of such superconductor with two halves occupied by the two states with opposite chiralities. Fractional flux plateaus are found in magnetization curve associated with free-energy minima, where the two domain walls between the two halves of loop accommodate different inter-component phase kinks leading to finite winding numbers around the loop only in a part of all condensates. Fractional flux plateaus form pairs related by the flux quantum $\Phi_0=hc/2e$, although they individually take arbitrary values depending on material parameters and temperature. This phenomenon is a clear evidence of TRS broken superconductivity, and in a general point of view it provides a novel chance to explore relative phase difference, phase kink and soliton in ubiquitous multi-component superconductivity such as that in iron pnicitides.
\end{abstract}

\date{\today}

\maketitle
\section{Introduction}
$\psi'^*_j$
Vortex with $2\pi$ phase winding is a hallmark of macroscopic
quantum state such as superfluidity and superconductivity
\cite{London50,Onsager61,Yang61,Abrikosov57}.
In superconductors, a vortex is accompanied by a quantum of magnetic flux  $\Phi_0=hc/2e$
in a closed path with zero supercurrent.
Since the quantization of magnetic flux is intimately related to the phase winding,
superconductivity gap functions carrying intrinsic phase variation induced by
unconventional pairing symmetry should leave unique consequences on the response to external magnetic field.
Several interesting examples are available.
A tricrystal ring of cuprate superconductor $\textrm{YBa}_2\textrm{Cu}_3\textrm{O}_{7-\delta}$ was observed
to carry a half flux quanta $\Phi_0/2$, which is the signature for
$d$-wave pairing symmetry \cite{Tsuei00}. In a ring-shaped setup composed of Nb and $\textrm{NdFeAsO}_{0.88}\textrm{F}_{0.12}$,
flux jumps in odd-number multiple of half flux quanta were observed \cite{Chen10,Chen09,Parker09}, which provide
support to the $S_{+-}$ pairing symmetry for iron-pnictide superconductors \cite{Mazin08, Kuroki08, Chen08, Graser09, Wang09, Seo08,Ishida09, Dai12}.
Half-valued fluxoid jumps in magnetization curve of a thin annular coil composed of $\textrm{Sr}_2\textrm{RuO}_4$ were
reported to be consistent with the $p$-wave pairing symmetry \cite{Jang11}.

\begin{figure}[t]
\psfig{figure=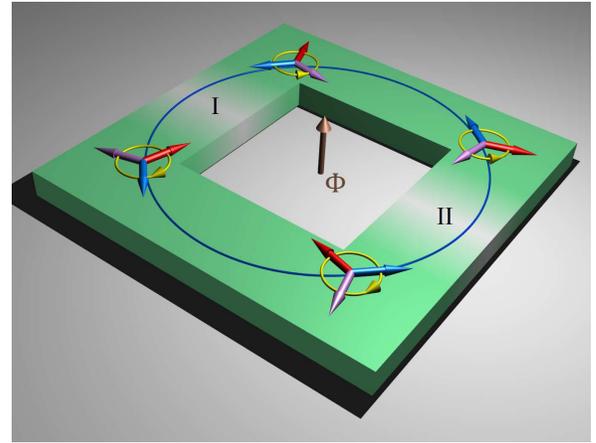,width=7.7cm} \caption{\label{f1} Schematic setup with a loop of time-reversal-symmetry-broken (TRSB) superconductor,
where the two halves are occupied by the two degenerate states with opposite chiralities. The three arrows denote the phases of order parameters and the yellow circle indicates the chirality
referring to the mutual phases ${\rm red}\rightarrow{\rm blue}\rightarrow{\rm purple}$
among three condensates. Between the two halves of the loop there are two domain wall I and II accommodating inter-component phase kinks.
}
\end{figure}

The degree of freedom of relative phase difference in a two-component superconductor was first discussed by Leggett \cite{Leggett66}, and experimental observations on the collective Leggett mode
were reported for MgB$_2$ \cite{Blumberg07,Nagamatsu01}.
Phase solitons in one of the two superconducting gap functions associated with fractional fluxoid jumps
have been investigated \cite{Sigrist99,Tanaka02,Babaev02,Gurevich03,Bluhm06,Vakaryuk12}.

Another interesting possibility in multi-component superconductors was raised some time ago \cite{Agterberg99} whereby Josephson-like inter-component
repulsions among three condensates can generate frustrations in phases of gap functions and induce a time-reversal-symmetry-broken (TRSB) superconducting state
even without external magnetic field, which corresponds to a high dimensional irreducible representation of point group similarly to the $p$-wave superconductor~\cite{Jang11}.
Due to the discovery of iron-pnictide superconductors where several orbitals of Fe contribute to multi
superconducting condensates, this possibility becomes realistic and a considerable amount of subsequent
works have been devoted to discuss its thermodynamic stability and various novel properties
\cite{Stanev10,Tanaka10,Carlstrom11,Dias11,Yanagisawa12,Hu12,Lin12,Maiti13,Wilson13,Orlova13,Marciani13,Takahashi14,Hinojosa14,Tanaka15}.

Recently it has been revealed by the present authors that, in a Josephson junction between a conventional single-component superconductor
and a multicomponent superconductor in the TRSB state, the critical current should be asymmetric with respect to the current direction as the consequence of
broken TRS \cite{Huang14}. As a matter of fact, unequal critical currents in opposite current directions were observed experimentally in a Josephson junction between $\textrm{PbIn}$ and $\textrm{BaFe}_{1.8}\textrm{Co}_{0.2}\textrm{As}_2$ \cite{Schmidt10}. Therefore, the TRSB state may have been realized already in iron-based superconductors,
which is consistent with a microscopic analysis where band structures and strongly correlated effects are taken into account \cite{Maiti13}.
Cross checking this novel superconducting phenomenon becomes an important issue.

In the present work, we address a new phase-sensitive property of the TRSB superconducting state. As schematically shown in Fig.~\ref{f1}, we
consider a loop of a multicomponent superconductor where the two halves are occupied by two TRSB states carrying opposite chiralities,
accompanied by two domain walls associated with inter-component phase kinks. We reveal explicitly that fractional flux plateaus appear in magnetization curve
associated with free-energy minima, where the domain walls accommodate phase kinks among different components leading to $2\pi$ phase winding along the loop
only in one or two of the three condensates. While the heights of fractional flux plateaus depend on material parameters and temperature,
they form pairs with heights related by the flux quantum $\Phi_0$, which is a unique signature of the TRSB superconducting state and can be used to confirm
the state itself. In a more general point of view this provides a novel chance to explore relative phase difference, phase kink and soliton in ubiquitous multicomponent superconductivity.

The remaining part of this paper is organized as follows. The mechanism for fractional flux plateaus in presence of domain walls is discussed in Sec.~II. We then simulate the magnetizing process with the time-dependent Ginzburg-Landau (TDGL) approach and confirm fractional flux plateaus associated with free-energy minima in Sec.~III. In Sec.~IV, we study an asymmetric loop where the widths of two domain walls differ. In Sec.~V we discuss the stability of the domain-wall structure and temperature dependence of the fractional flux plateaus. Finally we give a summary in Sec. VI.

\begin{figure}[t]
\psfig{figure=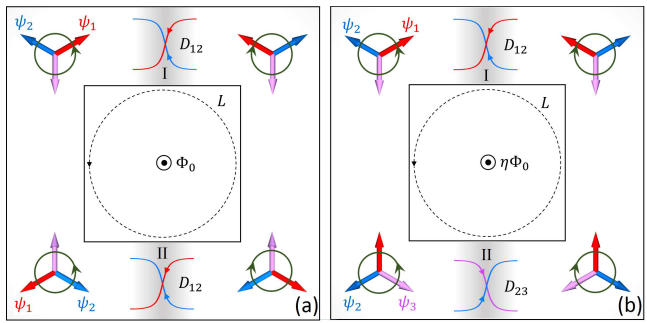,width=8.5cm} \caption{\label{f2} Illustration of relation between the
phase kinks at domain wall I and II and the flux trapped in the superconductor loop:
(a) phase kinks between a same pair of components which can only trap integer multiples of flux quantum;
(b) phase kinks between two different pairs of components which can trap a fractional flux ($0<\eta<1$).
Colored curves at the domain walls refer to the phase variations in corresponding condensates, and D$_{ij}$
is for the gauge-invariant phase kink between condensate $i$ and $j$.
The dashed circle denotes the direction for counting phase winding in the loop.}
\end{figure}

\section{Fractional flux in superconducting loop}
In order to reveal the essence of physics we first consider an isotropic TRSB state which is generated by three equivalent condensates with equal mutual repulsion.
For simplicity all order parameters are taken as $s$-wave from now on.
The two states in the loop are given by $\Psi=\{\psi_1,\psi_2,\psi_3\}=|\psi|\{1,e^{i2\pi/3},e^{i4\pi/3}\}$ and $\Psi^*$ with opposite chiralities (see Fig.~\ref{f1}).
Across each of the two domain walls between the left and right halves of the superconductor loop, there is a phase kink where the intercomponent
phase difference between two of the three order parameters shrinks to zero and reopens in the opposite way continuously, resulting in a sign reversal in the phase difference at the two sides of domain (see Fig.~2). We notice that phase kinks are gauge-invariant objects, which inevitably appear at the interface between two bulks of TRSB states with opposite chiralities.

\begin{figure}[t]
\psfig{figure=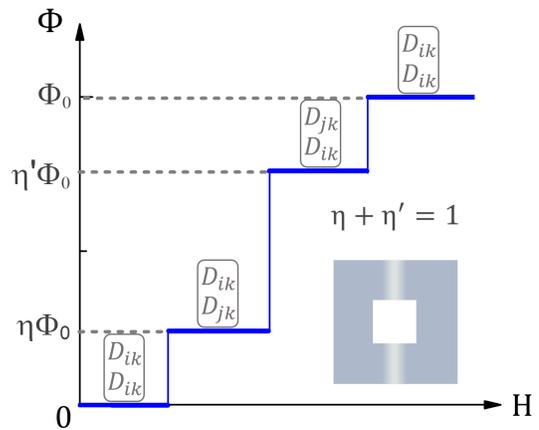,width=7cm} \caption{\label{f3} Schematic magnetization curve with fractional flux plateaus displayed together with gauge-invariant phase kinks
denoted by [D$_{ik}$/D$_{jk}]$ in the order of domain walls I and II.}
\end{figure}

When the two domain walls accommodate the phase kink between the same two condensates, such as that between condensate 1 and 2 defined as D$_{12}$ in Fig.~\ref{f2}(a), the
phase rotation integrated in a counterclockwise manner (indicated by $L$ in Fig.~\ref{f2}) over the two domain walls
cancel each other, resulting in the same phase winding in all the three condensates. In this case, the flux trapped in the loop is an integer multiple of flux quantum $\Phi_0$ when the loop is thick enough to fully screen the magnetic field.

The situation differs when the two domain walls accommodate different phase kinks, such as D$_{12}$ and D$_{23}$ in domain wall I and II respectively shown
in Fig.~\ref{f2}(b). By inspection one sees that $\psi_2$ rotates $4\pi/3$ anticlockwise over the two domain walls, while $\psi_1$ and $\psi_3$ rotate $-2\pi/3$. When the external magnetic field provides the additional phase rotation of $2\pi/3$ in all condensates,
a state with $2\pi$ phase winding in $\psi_2$ and $0$ in both $\psi_1$ and $\psi_3$ is stabilized. This yields a fractional flux quanta $\Phi_0/3$ in the loop. The state with a fractional flux trapped in this loop is expected to be stable in a certain range of external magnetic field, which leads to a fractional flux plateau in magnetization curve shown schematically in Fig.~\ref{f3}.

The above discussion can be elucidated by the integration of magnetic flux over the superconductor loop using the GL formalism where the supercurrent is given by \cite{Tinkhambook,Gurevich07}
\begin{equation}\label{supercurrent}
{\bf J}=\sum_{j=1,2,3}\frac{2e}{m_j}|\psi_j|^2 \hbar\left(\nabla\varphi_j-\frac{2\pi}{\Phi_0}{\bf A}\right),
\end{equation}
 with $m_j$ and $\varphi_j$ the effective mass and phase of component-$j$. For a thick loop, the supercurrent is zero deep inside the superconductor. In this case the magnetic flux
trapped in the loop is given by the line integration of phase differences as can be seen from Eq.~(\ref{supercurrent})
\begin{eqnarray}\label{flux expression}
\Phi=\frac{\Phi_0}{2\pi}\left[\oint_{C}\frac{p_1\nabla\varphi_{1}+p_2\nabla\varphi_{2}+p_3\nabla\varphi_{3}}{p_1+p_2+p_3}dl\right]\ \ \ \ \ \nonumber \\
=\frac{\Phi_0}{2\pi}\left[\oint_C\nabla\varphi_1 dl +\int_{\rm DW}\frac{p_2\nabla\varphi_{12}+p_3\nabla\varphi_{13}}{p_1+p_2+p_3}dl\right],
\end{eqnarray}
with $p_j=|\psi_j|^2/m_j$ and $\varphi_{ij}=\varphi_j-\varphi_i$ for $i,j=1,2,3$, where \textit{"C"} is a closed path along the loop with zero supercurrent everywhere (see Fig.~1), and the "DW" denotes domain-wall regimes (gray parts in Fig.~2) with phase kinks. In the second line, we divide the integrand into two terms, indicating two contributions to the total magnetic flux. The first contribution should be an integer multiple of $2\pi$ due to the single-valued wave function in the loop. The integrand in the second contribution is nonzero only on domain walls. This contribution is nonzero when two different phase kinks are realized at domain wall I and II, with the value
depending also on the quantities $p_j$.

Presuming the same length of domain walls I and II, the
two configurations [D$_{ik}$/D$_{jk}$] and [D$_{jk}$/D$_{ik}$] at the domain walls [I/II] take
the same free energy. However, integrating phase differences for these two configurations
along the closed path in the counterclockwise manner (see Fig.~\ref{f2}) results in opposite fractional values of $2\pi$ in the
second term in Eq.~(\ref{flux expression}). Therefore, these two configurations give two
fractional fluxes $\Phi_1$ and $\Phi_2$ related by the flux quantum $\Phi_0$. Fractional flux plateaus with corresponding
configurations of phase kinks are schematically shown in Fig.~\ref{f3}.

\hspace{2cm}
\section{TDGL approach}
Here we adopt the GL formalism to check the thermodynamic stability of states carrying fractional fluxes. The GL free-energy functional of a three-band superconductor with Josephson-like inter-component couplings is given by \cite{Gurevich07,Hu12}
\begin{eqnarray}\label{free energy}
\hspace{0.0cm}
F&=\sum\limits_{j=1,2,3}{\left[{\alpha_j}\left|\psi_j\right|^2+\frac{\beta_j}{2}\left|\psi_j\right|^4+\frac{\hbar^2}{2m_j}\left|\left(\frac{\nabla}{i}-\frac{2\pi}{\Phi_0}{\bf{A}}\right)\psi_j\right|^2\right]} \nonumber \\
\hspace{-0.6cm}
&-\sum\limits_{j,k=1,2,3;j<k}\gamma_{jk}\left(\psi_j^*\psi_k+c.c.\right)+\frac{1}{8\pi}\left(\nabla\times \bf{A}\right)^2, \label{eq: a}
\end{eqnarray}
where $\alpha_j$ is a temperature-dependent coefficient which is negative when $T<T_{cj}$ and positive when $T>T_{cj}$, with $T_{cj}$ the
critical point of the superconducting component-$j$ before considering intercomponent couplings, and $\gamma_{jk}$ is
an intercomponent coupling taken as constant for simplicity. For \(\gamma_{12}\gamma_{13}\gamma_{23}<0\), a TRSB superconducting state
appears when the
coefficients in Eq.~(\ref{free energy}) satisfy conditions revealed in a previous work~\cite{Hu12}.
To be specific, we put \(\gamma_{12}, \gamma_{13}, \gamma_{23}<0\), namely all repulsive Josephson-like couplings, since it is easy to see
that other TRSB states can be generated from this case by a simple gauge transformation.

\begin{figure}[t]
\psfig{figure=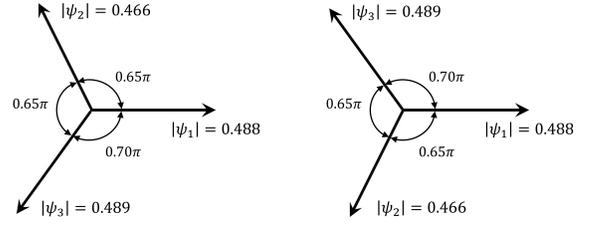,width=7.5cm} \caption{\label{f4} Amplitudes and phases of order parameters in two TRSB states with parameters $\alpha'_1=0.012, \alpha'_2=0.013$, $\alpha'_3=0.011$, $\gamma'_{12}=\gamma'_{23}=-0.24$, $\gamma'_{13}=-0.25$, $m'_1=m'_3=1$, $m'_2=1.1$, $\beta'_1=\beta'_2=\beta'_3=1$ and $\kappa_1=1.5$. See text for definitions of the dimensionless
GL parameters.
 }
\end{figure}

We adopt dimensionless quantities given by \cite{Lin11}
\begin{eqnarray}
&{\bf x}=\lambda_1(0){\bf x}',\ {\bf A}=\lambda_1(0)H_{tc1}(0)\sqrt{2}{\bf A}',\ {\bf J}=\frac{2e\hbar\psi_{10}^2(0)}{m_1\xi_1(0)}{\bf J'}, \nonumber \\
&\psi_j=\psi_{10}(0)\psi_j',\ \alpha_j=-\alpha_1(0)\alpha'_j, \beta_j=\beta_1\beta'_j,\ \gamma_{jk} =-\alpha_1(0)\gamma_{jk}',\nonumber \\
&m_j=m_1m_j{'},\ \kappa_1=\lambda_1(0)/\xi_1(0),\ F=G_0F' \nonumber
\end{eqnarray}
with $\psi^2_{10}(0)=-\alpha_1(0)/\beta_1$, $\lambda_1(0)=\sqrt{m_1c^2/[4\pi\psi^2_{10}(0)(2e)^{2}}]$, $\xi_1(0)=\sqrt{-\hbar^2/[2m_1\alpha_1(0)}]$, $H_{tc1}=\sqrt{-4\pi\alpha_1(0)\psi^2_{10}(0)}$ and $G_0=H_{tc1}^2(0)/4\pi$.
In the dimensionless units, the GL free energy is rewritten as
\begin{eqnarray}\label{dimensionless}
\hspace{0.0cm}
F'=&\sum\limits_{j=1,2,3}{\left[\alpha_j'\left|\psi_j'\right|^2+\frac{\beta_j'}{2}\left|\psi'_j\right|^4+\frac{1}{m_j{'}}\left|\left(\frac{1}{i\kappa_1}
\nabla-{\bf{A'}}\right)\psi_j'\right|^2\right]} \nonumber \\
&-\sum\limits_{j,k=1,2,3;j<k}\gamma_{jk}'\left(\psi_j^*{'}\psi_k'+c.c.\right)+\left(\nabla\times \bf{A'}\right)^2.
\end{eqnarray}

The system can be described by the following TDGL equations in the zero-electric potential gauge~\cite{Gropp96}
\begin{equation}\label{TDGL1}
\frac{\partial\psi_j'}{\partial{t}}=-\alpha_j'\psi_j'-\beta_j'|\psi_j'|^2\psi_j'-\frac{1}{m_j{'}}\left(\frac{1}{i\kappa_1}\nabla-{\bf A'}\right)^2\psi_j'+
\sum\limits_{k=1,2,3;k\neq j}\gamma_{jk}'\psi_k'
\end{equation}
with $j=1,2,3$ and
\begin{equation}\label{TDGL2}
{\sigma}\frac{\partial\bf{A'}}{\partial{t}}=\sum_{j=1,2,3}\frac{1}{m_j'}|\psi_j'|^2\left(\frac{1}{i\kappa_1}\nabla\varphi_j'-{\bf A'}\right)-\nabla\times\nabla\times{\bf A'}
\end{equation}
with $\sigma$ the coefficient of normal conductivity.
At equilibrium the left-hand sides of Eqs.~(\ref{TDGL1}) and (\ref{TDGL2}) are zero, which gives the GL equations. By solving three GL equations in Eq.~(\ref{TDGL1}) with ${\bf A'}=0$, we obtain the amplitudes and phases of the condensates at zero magnetic field~\cite{Hu12}.

\begin{figure}[t]
\psfig{figure=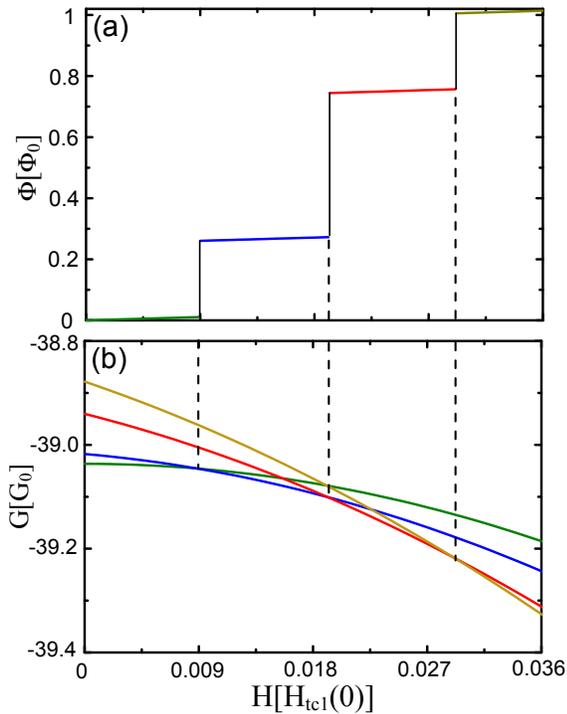,width=7.5cm} \caption{\label{f5}(a) Magnetization curve of thermodynamically stable state and (b) Gibbs free energies of several competing states in the superconductor
loop shown in Fig.~1 upon sweeping external magnetic field.
Parameters are the same as Fig.~\ref{f4}. The superconductor loop is of square shape with outer frame of
$24\lambda_1(0)$ and inner frame of $8\lambda_1(0)$.
}
\end{figure}

In the present multi-component superconducting system, the critical point $T_c$ is higher than $T_{cj}$ for $j=1,2,3$ \cite{Hu12}. In order to make sure that the GL formalism
is valid for investigating the thermodynamics properties of the coupled system, we choose temperature satisfying $T_{cj}<T\lesssim T_c$, where $\alpha_j'$ are positive and small. In Fig.~\ref{f4} we display the two TRSB states with opposite chiralities, which are used for the following study on the magnetization process.

We take a square loop with external dimension $24\lambda_1(0)\times 24\lambda_1(0)$ and the wall thickness $8\lambda_1(0)$ to investigate the magnetic response.
From Eq.~(\ref{supercurrent}) we obtain the the penetration length $\lambda^2=(\sum_j|\psi_j'|^2/m_j')^{-1}\lambda^2_1(0)$ in the dimensionless form. For the states given in Fig.~\ref{f4}, $\lambda=1.22\lambda_{1}(0)$ is much smaller than the thickness of loop wall. Therefore, one can take the closed path "C" as the middle line of the superconductor loop where supercurrent is negligibly small (see the Appendix). At the edge of superconductor, we presume that no supercurrent flows out of the superconductor,
and the ${\bf B}$ field at the external edge of superconductor loop is fixed to the value of applied magnetic field.

The magnetization curve derived from TDGL equations (\ref{TDGL1}) and (\ref{TDGL2}) is shown in Fig.~\ref{f5}, where fractional flux plateaus corresponding to states with free-energy minima are obtained.
When the magnetic field is small, there is no flux penetrating into the loop.
 As the magnetic field increases to $H/H_{tc1}(0)=0.009$, the stable state takes a {\it fractional flux} $\Phi_1=0.26\Phi_0$. This state remains stable until the magnetic field
$H/H_{tc1}(0)=0.019$, yielding a plateau in the magnetization curve. As the magnetic field increases further,
another state appears with fractional flux $\Phi_2=0.74\Phi_0$ in the regime
$0.019\leq H/H_{tc1}(0)\leq0.029$. We can see that both $\Phi_1$ and $\Phi_2$ deviate from $\Phi_0/3$ and $2\Phi_0/3$, because of the three inequivalent condensates in the system.
Nevertheless, it is clear that the relation $\Phi_1+\Phi_2=\Phi_0$ is satisfied as revealed in Sec.~II.
For even larger magnetic fields, the stable state permits one flux quantum $\Phi_0$
inside the loop. Note that the magnetic fields for the fractional flux plateaus are very small as compared with the typical field
$H_{tc1}(0)$ and thus there is no vortex inside the body of the superconductor.

We check the phase kinks at the two domain walls I and II and the phase windings along the loop
in the three condensates at the fractional flux plateaus.
At $\Phi_1=0.26\Phi_0$, phase kinks $D_{12}$ and $D_{23}$ are realized at regions I and II
respectively, and $\psi_2$ rotates $2\pi$ along the loop leaving $\psi_1$ and $\psi_3$ unwinding.
At $\Phi_1=0.74\Phi_0$, the phase kinks
$D_{12}$ and $D_{23}$ are at domain walls II and I, in contrast to the case of $\Phi_1=0.26\Phi_0$, and
$\psi_1$ and $\psi_3$ rotate $2\pi$ with $\psi_2$ unwinding.
At integer flux quanta $\Phi=0$ and $\Phi=\Phi_0$, the phase kink is $D_{12}$
at both domain walls I and II. All these are in accordance with the discussion in Sec.~II.
In general, there are at most six fractional flux plateaus between the integer flux quanta 0 and $\Phi_0$. For the GL parameters given in Fig.~\ref{f4} we can only see two plateaus in Fig. \ref{f5}(a) because the free energies of domain walls satisfy $F(D_{12})\lesssim F(D_{23})<F(D_{13})$ such that only the phase-kink pair D$_{12}$ and D$_{23}$ is stabilized.

\begin{figure}[t]
\psfig{figure=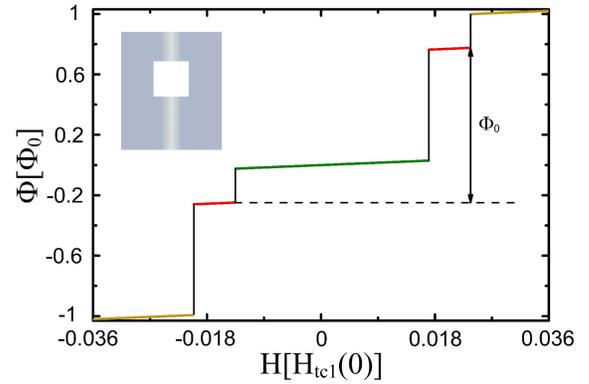,width=7.5cm} \caption{\label{f6} {(color online)} Magnetization curve with two fractional flux plateaus
for an asymmetric superconductor loop, where the width of side including domain II is enlarged to $12\lambda_1(0)$ from that given in Fig.~\ref{f5}. The parameters are the same as Fig.~\ref{f4} except for $\alpha_1'=0.025$, $\alpha_2'=0.028$ and $\alpha_3'=0.022$.
}
\end{figure}

\section{Asymmetric superconductor loop}
Up to this point, the superconductor loop is presumed to have the same width at domain wall I and II, which guarantees the degeneracy of domain-wall energy between
phase-kink pairs [D$_{ik}$/D$_{jk}$] and [D$_{jk}$/D$_{ik}$]. In this case, it is easy to see that the
magnetization curve in Fig.~\ref{f3} should be symmetric with respect to the direction of magnetic field. In general, the two widths can be different.
In the latter case [D$_{ik}$/D$_{jk}$] may be unstable even though [D$_{jk}$/D$_{ik}$] is stable and associated with the free-energy minimum.
As shown in Fig.~\ref{f6} for an asymmetric superconductor loop, the fractional flux plateau at $\Phi=0.74\Phi_0$ remains stable for
$0.017\leq H/H_{tc1}(0)\leq0.023$, while that at $\Phi=0.26\Phi_0$
disappears in contrast to Fig.~\ref{f5}, since they are associated with different phase-kink configurations at domain wall I and II.
It is worth noticing that even in this asymmetric loop the plateau at $\Phi=-0.26\Phi_0$ is still stable for
$-0.020\leq H/H_{tc1}(0)\leq -0.014$, since it is
associated with the same phase-kink configuration with that at $\Phi=0.74\Phi_0$ and a difference of flux quantum $\Phi_0$ coming from the first term in Eq.~(\ref{flux expression}).
In this asymmetric loop, the magnetization curve is asymmetric with respect to the direction of the magnetic field as in Fig.~\ref{f6}.
The property that fractional flux plateaus in positive and
negative magnetic fields are paired with the difference of flux quantum $\Phi_0$ is robust, and
can be taken as a crosscheck for fractional flux plateaus originated from the TRSB state.

\begin{figure}
\psfig{figure=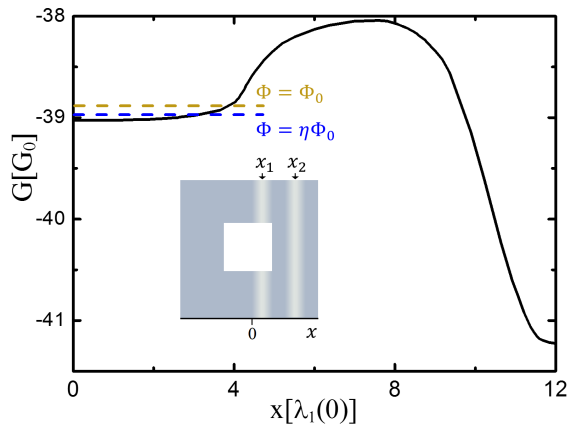,width=7.5cm} \caption{\label{f7} {(color online)} Gibbs free energy for
the system in Fig.~1 with two halves occupied by the two TRSB states with opposite chiralities
as a function of the location of domain walls $x$ defined in the inset.
Parameters and sample size are the same as Fig.~5.
}
\end{figure}

\section{Discussions}
In the present work we study the case that the left and right halves of the superconductor loop take the
two TRSB states with opposite chiralities. This situation can be realized in experiments by cooling the whole system from temperature above $T_{\rm c}$ with
laser heat pulse irradiated on regions I and II \cite{Tate89}. The two halves condensate independently and by chance arrive at the different TRSB superconducting states,
leading to the two domain walls at region I and II after releasing the irradiation. In order to check the stability of
this configuration, we estimate the free energy of the whole system in terms of the TDGL approach. As shown in Fig.~\ref{f7},
the state with two domain walls located at the middle of the top and bottom sides of the loop corresponds to
a free-energy minimum. The domain walls once generated should be stable since moving them outside the loop is prohibited by a large free-energy barrier
which is produced by an elongated, single domain wall during the process
of domain-wall relocation (see that at $x_2$ in the inset of Fig.~\ref{f7}). The stability of the present
setup against relocating one of the two domain walls along the loop can be provided by widening the left and right arms of the loop.
The increase in free energy in states with fractional fluxes and integer
flux quanta upon application of external magnetic field is smaller by one order of magnitude than the free-energy
barrier as seen in Fig.~\ref{f7}, which justifies the discussion on fractional flux plateaus in the present work.

At this point we notice that the free-energy barrier in Fig.~7 generated by the two TRSB states at the
two halves of superconductor loop is crucially important for the thermodynamic
stability of fractional flux plateaus. From Eq.~(\ref{flux expression}) one might think that a two-component
superconductor or a three-component one with preserved TRS can also accommodate fractional fluxes.
However, in these cases there is no free-energy barrier like that in Fig.~7, and states with fractional
fluxes are unstable.

The amplitudes and inter-component phase differences of order parameters
change with temperature, leading to variation in stable domain-wall structures. As a result, both height and width
of a fractional flux plateau should change as temperature is swept. During field-cooling or field-heating processes,
jumps between states with fractional fluxes and/or integer multiples of flux quantum can take place.

Half-valued fluxoid jumps in magnetization curve were reported in a small multicomponent superconducting sample comparable with penetration length\cite{Jang11}. This results in incomplete screening of magnetic field, and thus the magnetization curve exhibits a finite slope for any external magnetic field. Namely, there is no fractional flux plateau in their system. In previous studies on vortex states of TRSB superconductor it was discussed that vortex cores of different condensates can deviate from each other in space \cite{Garaud11,Takahashi14,Garaud14,Gillis14,HuangPre}. However, without a closed path along which supercurrent is zero everywhere,
there is no fractional flux plateau in the magnetization curve.

\section{Conclusion} To summarize, we have studied the magnetic response of a loop of three-component superconductor with two degenerate time-reversal symmetry broken states at two halves.
When the two domain walls between the two halves accommodate different intercomponent phase kinks,
fractional flux plateaus appear in the magnetization curve which form pairs related to each other by the flux quantum.
These properties are expected to be helpful for detecting experimentally the time-reversal symmetry broken superconducting state
which can be realized in iron-pnictide superconductors. In general, this endeavour provides a novel chance to explore relative phase difference, phase
kink and soliton in ubiquitous multi-component superconductivity.

\section*{Acknowledgements} The authors are grateful to Masashi Tachiki, Zhi Wang, Yusuke Kato and Atsutaka Maeda for stimulating discussions. This work was supported by the WPI initiative on Materials Nanoarchitectonics, and partially by the Grant-in-Aid for Scientific Research (No. 25400385), MEXT of Japan.

\setcounter{figure}{0}
\renewcommand{\thefigure}{A\arabic{figure}}

\appendix
\section*{Appendix: Distribution of Supercurrent in the Superconductor Loop}

\begin{figure}
\psfig{figure=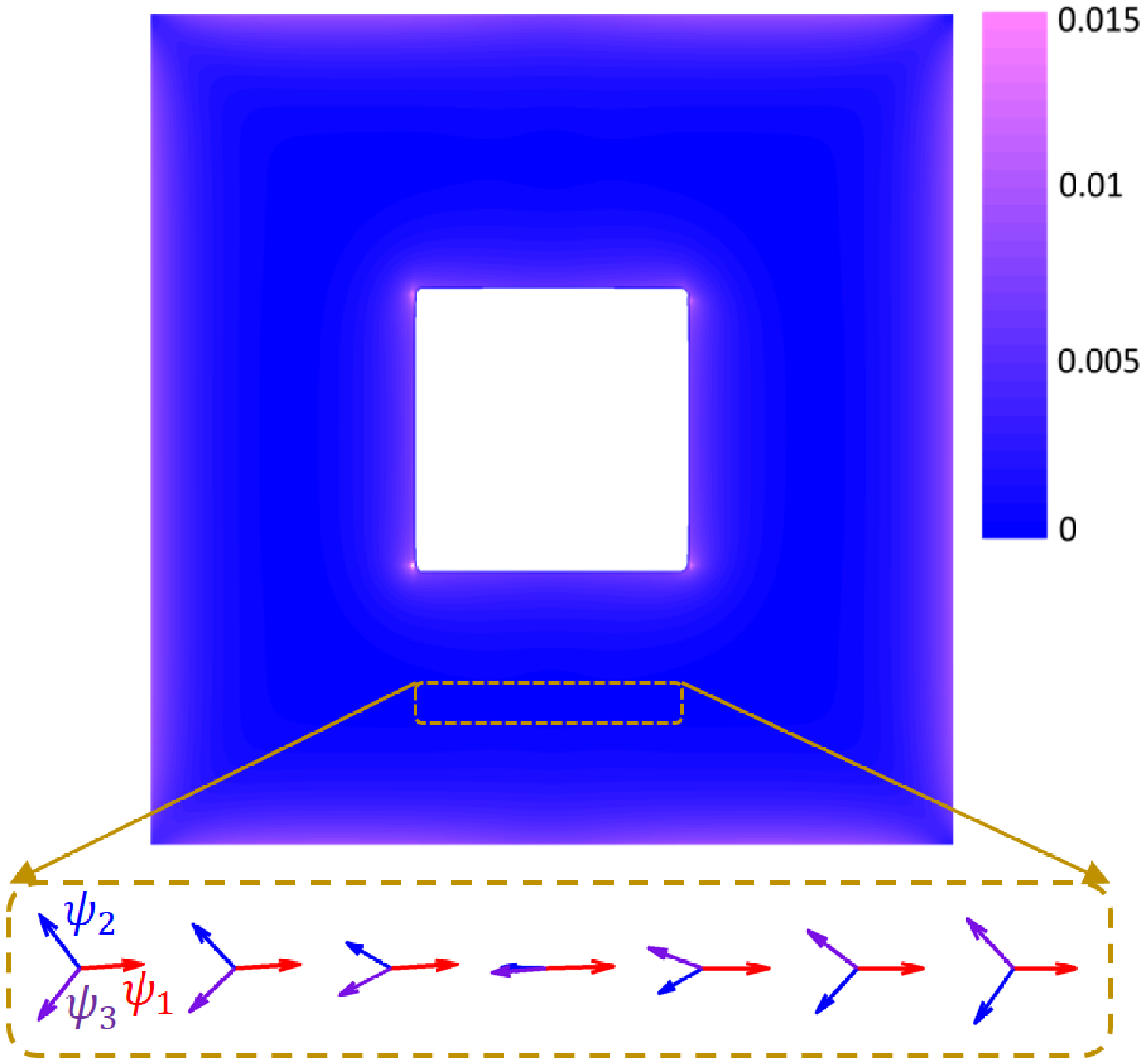,width=6cm} \caption{\label{f8} {(color online)} Distribution of supercurrent density ${\bf J'}$ in the superconductor loop for the fractional flux plateau of $\Phi=0.26\Phi_0$ in Fig.~5, and that of order parameters on domain wall II. The directions and lengths of arrows refer to phases and amplitudes of order parameters, where $|\psi_2|$ and $|\psi_3|$ are suppressed to $0.63$ and $0.77$ of the bulk values while $|\psi_1|$ remains almost unchanged.}
\end{figure}

An example of the distribution of supercurrent density in the superconductor loop and order parameters on domain walls is shown in Fig.~\ref{f8}, where the superconductivity survives in all components and the Meissner effect screens the magnetic field and thus the total supercurrent to zero. The important feature here is that, although the supercurrents in individual components are not zero due to the phase shifts, they flow in opposite directions and cancel each other, leading to zero total supercurrent. The situation of a domain wall as discussed in the present work differs considerably from a vortex, where the phase winding induces a divergent kinetic energy, which necessitates total suppression of the amplitude of the superconducting order parameter at the vortex core.

\begin{thebibliography}{34}%
\makeatletter
\providecommand \@ifxundefined [1]{%
 \@ifx{#1\undefined}
}%
\providecommand \@ifnum [1]{%
 \ifnum #1\expandafter \@firstoftwo
 \else \expandafter \@secondoftwo
 \fi
}%
\providecommand \@ifx [1]{%
 \ifx #1\expandafter \@firstoftwo
 \else \expandafter \@secondoftwo
 \fi
}%
\providecommand \natexlab [1]{#1}%
\providecommand \enquote  [1]{``#1''}%
\providecommand \bibnamefont  [1]{#1}%
\providecommand \bibfnamefont [1]{#1}%
\providecommand \citenamefont [1]{#1}%
\providecommand \href@noop [0]{\@secondoftwo}%
\providecommand \href [0]{\begingroup \@sanitize@url \@href}%
\providecommand \@href[1]{\@@startlink{#1}\@@href}%
\providecommand \@@href[1]{\endgroup#1\@@endlink}%
\providecommand \@sanitize@url [0]{\catcode `\\12\catcode `\$12\catcode
  `\&12\catcode `\#12\catcode `\^12\catcode `\_12\catcode `\%12\relax}%
\providecommand \@@startlink[1]{}%
\providecommand \@@endlink[0]{}%
\providecommand \url  [0]{\begingroup\@sanitize@url \@url }%
\providecommand \@url [1]{\endgroup\@href {#1}{\urlprefix }}%
\providecommand \urlprefix  [0]{URL }%
\providecommand \Eprint [0]{\href }%
\providecommand \doibase [0]{http://dx.doi.org/}%
\providecommand \selectlanguage [0]{\@gobble}%
\providecommand \bibinfo  [0]{\@secondoftwo}%
\providecommand \bibfield  [0]{\@secondoftwo}%
\providecommand \translation [1]{[#1]}%
\providecommand \BibitemOpen [0]{}%
\providecommand \bibitemStop [0]{}%
\providecommand \bibitemNoStop [0]{.\EOS\space}%
\providecommand \EOS [0]{\spacefactor3000\relax}%
\providecommand \BibitemShut  [1]{\csname bibitem#1\endcsname}%
\let\auto@bib@innerbib\@empty

\bibitem{London50} F.~London, {\it Superfluids}, Vol. I., Wiley, New York, (1950);
\bibitem{Onsager61} L.~Onsager, Phys.~Rev.~Lett. {\bf 7}, 50 (1961).
\bibitem{Yang61} N.~Byers and C.~N.~Yang, Phys.~Rev.~Lett. {\bf 7}, 46 (1961).
\bibitem{Abrikosov57} A.~A.~Abrikosov, Sov.~Phys.~JETP {\bf 5}, 1174 (1957).
\bibitem{Tsuei00} C.~C.~Tsuei and J.~R.~Kirtley, Rev.~Mod.~Phys. {\bf 72}, 969 (2000).
\bibitem{Chen10} C.~T.~Chen, C.~C.~Tsuei, M.~B.~Ketchen, Z.~A. Ren, and Z.~X.~Zhao, Nature~Phys. {\bf 6}, 260 (2010).
\bibitem{Chen09} W.~Q.~Chen, F.~Ma, Z.~Y.~Liu, and F.~C.~Zhang, Phys.~Rev.~Lett. {\bf 103}, 207001 (2009).
\bibitem{Parker09} D.~Parker and I.~I.~Mazin, Phys.~Rev.~Lett. {\bf 102}, 227007 (2009).
\bibitem{Mazin08} I.~I.~Mazin, D.~J.~Singh, M.~D.~Johannes, and M.~H.~Du, Phys.~Rev.~Lett. {\bf 101}, 057003 (2008).
\bibitem{Kuroki08} K.~Kuroki, S.~Onari, R.~Arita, H.~Usui, Y.~Tanaka, H.~Kontani, and H.~Aoki, Phys.~Rev.~Lett. {\bf 101}, 087004 (2008).
\bibitem{Chen08} W.~Q.~Chen, K.~Y.~Yang, Y.~Zhou and F.~C.~Zhang, Phys.~Rev.~Lett. {\bf 102}, 047006 (2008).
\bibitem{Graser09} S.~Graser, T.~A.~Maier, P.~J.~Hirschfeld, and D.~J.~Scalpino, New~J.~Phys. {\bf 11}, 025016 (2009).
\bibitem{Wang09} F.~Wang, H.~Zhai, Y.~Ran, A.~Vishwanath, and D.~H.~Lee, Phys.~Rev.~Lett. {\bf 102}, 047005 (2009).
\bibitem{Seo08} K.~Seo, B.~A.~Bernevig, and J.~Hu, Phys.~Rev.~Lett. {\bf 101}, 206404 (2008).
\bibitem{Ishida09} K.~Ishida, Y.~Nakai, and H.~Hosono, J.~Phys.~Soc.~Jpn. {\bf 78}, 062001 (2009).
\bibitem{Dai12} P.~C.~Dai, J.~P.~Hu, and E.~Dagotto, Nature~Phys. {\bf 8}, 709 (2012)
\bibitem{Jang11} J.~Jang, D.~G.~Ferguson, V.~Vakaryuk, R.~Budakian, S.~B.~Chung, P.~M.~Goldbart, and Y.~Maeno, Science {\bf 331}, 186 (2011).
\bibitem{Leggett66} A.~J.~Leggett, Prog.~Theor.~Phys. 36, 901 (1966).
\bibitem{Blumberg07} G.~Blumberg, A.~Mialitsin, B.~S.~Dennis, M.~V.~Klein, N.~D.~Zhigadlo, and J.~Karpinski, Phys.~Rev.~Lett. {\bf 99}, 227002 (2007).
\bibitem{Nagamatsu01} J.~Nagamatsu, N.~Nakagawa, T.~Muranaka, Y.~Zenitani, and J.~Akimitsu, Nature {\bf 410}, 63 (2001).
\bibitem{Sigrist99} M.~Sigrist and D.~F.~Agterberg, Prog.~Theor.~Phys. {\bf 102}, 965 (1999).
\bibitem{Tanaka02} Y.~Tanaka, Phys.~Rev.~Lett. {\bf 88}, 017002 (2002).
\bibitem{Babaev02} E.~Babaev, Phys.~Rev.~Lett. {\bf 89}, 067001 (2002).
\bibitem{Gurevich03} A.~Gurevich and V.~M.~Vinokur, Phys.~Rev.~Lett. {\bf 90} 047004 (2003).
\bibitem{Bluhm06} H.~Bluhm, N.~C.~Koshnick, M.~E.~Huber, and K.~A.~Moler, Phys.~Rev.~Lett. {\bf 97}, 237002 (2006).
\bibitem{Vakaryuk12} V.~Vakaryuk, V.~Stanev, W.~C.~Lee, and A.~Levchenko, Phys.~Rev.~Lett. {\bf 109}, 227003 (2012).
\bibitem{Agterberg99} D.~F.~Agterberg, V.~Barzykin, and L.~P.~Gor'kov, Phys.~Rev.~B {\bf 60}, 14868 (1999).
\bibitem{Stanev10} V.~Stanev and Z.~Tesanovic, Phys.~Rev.~B {\bf 81}, 134522 (2010).
\bibitem{Tanaka10} Y.~Tanaka and T.~Yanagisawa, J.~Phys.~Soc.~Jpn. {\bf 79}, 114706 (2010).
\bibitem{Carlstrom11} J.~Carlstrom, J.~Garaud, and E.~Babaev, Phys.~Rev.~B {\bf 84}, 134518 (2011).
\bibitem{Dias11} R.~G.~Dias and A.~M.~Marques, Supercond.~Sci.~Technol. {\bf 24}, 085009 (2011).
\bibitem{Yanagisawa12} T.~Yanagisawa, Y.~Takana, I.~Hase, and K.~Yamaji, J.~Phys.~Soc.~Jpn. {\bf 81}, 024712 (2012).
\bibitem{Hu12} X.~Hu and Z.~Wang, Phys.~Rev.~B {\bf 85}, 064516 (2012).
\bibitem{Lin12} S.~Z.~Lin and X.~Hu, Phys.~Rev.~Lett. {\bf 108}, 177005 (2012).
\bibitem{Maiti13} S.~Maiti and A.~V.~Chubukov, Phys.~Rev.~B {\bf 87}, 144511 (2013).
\bibitem{Wilson13} B.~J.~Wilson and M.~P.~Das, J.~Phys.:~Condens.~Matter {\bf 25}, 425702 (2013).
\bibitem{Orlova13} N.~V.~Orlova, A.~A.~Shanenko, M.~V.~Milosevic, and F.~M.~Peeters, Phys.~Rev.~B {\bf 87}, 134510 (2013).
\bibitem{Marciani13} M.~Marciani, L. Fanfarillo, C. Castellani, and L. Benfatto, Phys.~Rev.~B {\bf 88}, 214508 (2013).
\bibitem{Takahashi14} Y.~Takahashi, Z.~Huang, and X.~Hu, J.~Phys.~Soc.~Jpn. {\bf 83}, 034701 (2014).
\bibitem{Hinojosa14} A.~Hinojosa, R.~M.~Fernandes, and A.~V.~Chubukov, Phys.~Rev.~Lett. {\bf 113}, 167001 (2014).
\bibitem{Tanaka15} Y.~Tanaka, Supercond.~Sci.~Technol. {\bf 28}, 034002 (2015).
\bibitem{Huang14} Z.~Huang and X.~Hu, Appl.~Phys.~Lett. {\bf 104}, 162602 (2014).
\bibitem{Schmidt10} S.~Schmidt, S.~Doring, F.~Schmidl, V.~Grosse, and P.~Seidel, Appl.~Phys.~Lett. {\bf 97}, 172504 (2010).
\bibitem{Tinkhambook} M. Tinkham, \emph{Introduction to Superconductivity} (McGraw-Hill, Inc., New York, 1996).
\bibitem{Gurevich07} A.~Gurevich, Physica~C {\bf 456}, 160 (2007).
\bibitem{Lin11} S.~Z.~Lin and X.~Hu, Phys.~Rev.~B {\bf 84}, 214505 (2011).
\bibitem{Gropp96} W.~D.~Gropp, H.~G.~Kaper, G.~K.~Leaf, D.~M.~Levine, M.~Palumbo, and V.~M.~Vinokur, J.~Comput.~Phys. {\bf 123}, 254 (1996).
\bibitem{Tate89} J.~Tate, B.~Cabrera, S.~B.~Felch, and J.~T.~Anderson, Phys.~Rev.~Lett. {\bf 62}, 845 (1989).
\bibitem{Garaud11} J.~Garaud, J.~Carlstrom, and E.~Babaev, Phys.~Rev.~Lett. {\bf 107}, 197001 (2011).
\bibitem{Garaud14}J.~Garaud and E.~Babaev, Phys.~Rev.~Lett. 112, 017003 (2014).
\bibitem{Gillis14} S.~Gillis, J.~Jaykka, and M.~V.~Milosevic, Phys.~Rev.~B {\bf 89} 024512 (2014).
\bibitem{HuangPre} Z.~Huang and X.~Hu, J.~Supercond.~Nov.~Magn. (in~press,~arXiv:1510.05890).


\end{thebibliography}
%

\end{document}